\documentclass[aps,prmaterials,reprint,groupedaddress,longbibliography]{revtex4-2}
\usepackage{graphicx,longtable,float,textcomp}

\begin{document}


\title{\textit{Ab initio} construction of full phase diagram of MgO-CaO eutectic system using neural network interatomic potentials}


\author{Kyeongpung Lee}
\author{Yutack Park}
\author{Seungwu Han}
\email{hansw@snu.ac.kr}
\affiliation{Department of Materials Science and Engineering and Research Institute of Advanced Materials, Seoul National University, Seoul 08826, Korea}


\date{\today}

\begin{abstract}

While several studies confirmed that machine-learned potentials (MLPs) can provide accurate free energies for determining phase stabilities, the abilities of MLPs for efficiently constructing a full phase diagram of multi-component systems are yet to be established. 
In this work, by employing neural network interatomic potentials (NNPs), we demonstrate construction of the MgO-CaO eutectic phase diagram with temperatures up to 3400 K, which includes liquid phases. The NNP is trained over trajectories of various solid and liquid phases at several compositions that are calculated within the density functional theory (DFT). For the exchange-correlation energy among electrons, we compare the PBE and SCAN functionals. The phase boundaries such as solidus, solvus, and liquidus are determined by free-energy calculations based on the thermodynamic integration or semigrand ensemble methods, and salient features in the phase diagram such as solubility limit and eutectic points are well reproduced. In particular, the phase diagram produced by the SCAN-NNP closely follows the experimental data, exhibiting both eutectic composition and temperature within the measurements. On a rough estimate, the whole procedure is more than 1,000 times faster than pure-DFT based approaches. We believe that this work paves the way to fully \textit{ab initio} calculation of phase diagrams.
\end{abstract}

\maketitle

\section{INTRODUCTION}
By informing phase formation under the given temperature, pressure, or composition, the phase diagram plays an important role in designing and processing materials~\cite{intro1,intro2}. 
However, determination of the phase diagram requires a huge amount of experimental efforts, particularly for multicomponent systems~\cite{new11,mgocao10}. This is because while possible combinations of temperature and composition are vast, each data point becomes only reliable with consistent observations from complementary techniques. As such, full phase diagrams are sparse for multicomponent systems~\cite{new11}.

Theoretically, the phase diagram is determined by the Gibbs free energies of competing phases, where the lowest ones appear in the equilibrium phase diagram. Several computational methods based on molecular dynamics (MD) have been developed for computing the free energies from atomistic simulations: thermodynamic integration, coexistence method, and semigrand ensemble~\cite{ti1,ti2,ip1,sgc0}. In combination with the density functional theory (DFT), these methods allow for evaluating free energies without experimental inputs. For example, various single-component phase diagrams including melting properties have been constructed by employing the above-mentioned methods~\cite{dftu1,dftu2,dftu3,dftu4,dftu5,dftu6,dftu7}. However, in binary or higher-order systems, the MD-based approaches are limited with DFT because the sampling over compositional variations and configurations requires iterative simulations over millions of time steps and large simulation cells containing hundreds of atoms~\cite{sgc1,sgc2}. Alternatively, the MD-free cluster expansion was employed in constructing phase diagrams of solid alloys by interpolating free energies of alloy configurations~\cite{ce1,ce2,ce3,ce4,ce5,ce6,dftb1,dftb2}. However, this approach is applicable to only crystal systems, and its accuracy degrades when atomic relaxations are significant~\cite{ce7}.

In recent years, machine-learned potentials (MLPs) have gained much attention as they can provide energies with near-DFT accuracy at a fraction of the cost~\cite{mlpgen1}. The computational acceleration using MLPs has been confirmed over a wide range of applications including, for example, crystal structure prediction~\cite{csp1} and lattice thermal conductivity~\cite{ltc1}. In addition, MLPs are suitable as surrogate models of DFT in evaluating free energies, which has been successfully demonstrated in many recent studies.~\cite{mlpu1,mlpu2,mlpu3,mlpu4,mlpu5,mlpu6,mlpu7,mlpu8,mgocao6,mlpadv1,mlpadv2,mlpb3,mlpb1,mlpb2}
However, examples are mostly single-component systems~\cite{mlpu1,mlpu2,mlpu3,mlpu4,mlpu5,mlpu6,mlpu7,mlpu8,mgocao6,mlpadv1,mlpadv2} and only a few examples, Ag$_{x}$Pd$_{1-x}$~\cite{mlpb3}, Ni$_x$Mo$_{1-x}$~\cite{mlpb1}, and Ga$_x$As$_{1-x}$~\cite{mlpb2}, have been attempted for constructing the phase diagram of compounds. Therefore, the accuracy and efficiency of MLPs for constructing the whole phase diagram of multi-component systems are yet to be confirmed. 
With these motivations, herein we aim to construct a full temperature-composition phase diagram for the MgO-CaO, an archetypal pseudo-binary system with rich experimental information, using Behler-Parrinello-type neural network potentials (NNPs)~\cite{bp1}. 

Our strategy for computing the free energy and constructing the phase diagram is as follows: first, for pure phases, temperature-dependent free energies are calculated using the thermodynamic integration method. For pure MgO and CaO, we consider rocksalt and liquid phases, and the crossing of the free energy curves of both phases corresponds to the melting point. Next, the composition-dependent free energy of mixing is calculated using semigrand ensemble simulations at selected temperatures. Since no intermetallic compound exists along the MgO-CaO pseudobinary line, only the rocksalt solid solution phase and liquid mixture are considered. The whole temperature- and composition-dependent free energies are fitted into analytical forms, and phase boundaries are determined by common tangents on the isothermal sections of free energy curves. 
The rest of the paper is organized as follows: in Sec.~\ref{secmethod} we introduce computational methods used in the present work such as NNPs, thermodynamic integration, and semigrand ensemble simulations. The main results are discussed in Sec.~\ref{secresult}, and Sec.~\ref{secsummary} summarizes and concludes this work.

\section{THEORETICAL METHODS}
\label{secmethod}
\subsection{Neural network potential and DFT calculations}
In the present work, the NNPs are trained by using the \texttt{SIMPLE-NN} package~\cite{simplenn1,simplenn2}. For input features, we use atom-centered symmetry functions (ACSFs)~\cite{bp2}. The numbers of features are 24 and 108 for the radial and angular parts, respectively, with cutoff radii of 7.0 and 4.5 \AA, respectively. The full parameters for ACSFs are listed in the Supplementary Material. The training is accelerated by decorrelating features using principal component analysis and whitening~\cite{pca1}. We use an initial learning rate of 0.01, which decays exponentially during 190 epochs and becomes 0.0005 at the final epoch. We use a fully connected atomic neural network with two 60-node hidden layers. The MD simulations and evaluations of energy, force, and stress are carried out using the \texttt{LAMMPS} package~\cite{simplenn1,lammps1}.

The DFT calculations for the training set are carried out using Vienna $Ab$ $initio$ Simulation Package (\texttt{VASP})~\cite{vasp1,vasp2,vasp3} with the projector-augmented wave  pseudopotentials~\cite{vasp4}. The pseudopotential contains the valence electrons of $3s^2$, $3s^23p^64s^2$, and $2s^22p^4$ for Mg, Ca, and O, respectively. We generate data sets independently using two types of the exchange-correlation functional; the widely-used the generalized gradient approximation (GGA) by Perdew-Burke-Ernzerhof (PBE)~\cite{pbe1} and strongly constrained and appropriately normed (SCAN) meta-GGA functional~\cite{scan1}. 
The SCAN functional has been benchmarked against PBE on diverse properties, providing more accurate lattice parameters~\cite{scan1}, formation enthalpies~\cite{scan2}, lattice dynamics~\cite{scan4}, energies of metastable phases~\cite{scan5}, and the melting points~\cite{mgocao5,mgocao6}. For \textit{ab initio} MD simulations, we use default plane-wave energy cutoffs with the $\Gamma$-point sampling for the Brillouin zone integration. Then more accurate DFT calculations are performed on selected snapshots for reference data set by increasing the plane-wave energy cutoff to 500 eV and employing 3$\times$3$\times$3 {\bf k}-point meshes for the conventional unit cell of rocksalt MgO and CaO, which is scaled in supercells to select a similar {\bf k}-point density. Details on the training structures will be discussed in Sec.~\ref{sec3a}.

\subsection{Thermodynamic integration}
Thermodynamic integration allows one to calculate the free energy by computing the work done in the isothermal switching process from a reference state whose free energy is known \textit{a priori}, to a state of interest~\cite{ti1,ti2,ti3,dftu4}. We apply this method for pure rocksalt and liquid phases of MgO and CaO. When the potential energy term of Hamiltonian of the reference system ($U_{\rm i}$) and of the system of interest ($U_{\rm f}$) is given, a parametric potential is defined as
\begin{equation}
	U(\lambda) = (1-\lambda) U_{\rm i}+{\lambda}U_{\rm f} ,
\end{equation}
where $\lambda$ is a coupling parameter ranging from 0 to 1. The difference in the Helmholtz free energy between the two systems ($F_{\rm f}-F_{\rm i}$) is given by
\begin{equation}
	\label{eq_ti} F_{\rm f}-F_{\rm i} = \int_{0}^{1}{\left\langle\frac{\partial U(\lambda)}{\partial \lambda}\right\rangle_{\lambda}d\lambda} ,
\end{equation}
where the $\left\langle..\right\rangle_{\lambda}$ denotes the ensemble average under the NVT condition at constant $\lambda$, which is practically replaced by a temporal average according to the ergodicity.

We employ two reference systems depending on the final state: the Einstein crystal for solid phases and Lennard-Jones (LJ) fluid for liquid phases. The free energy of Einstein crystal is given by
\begin{equation}
	F = \sum_{i}{3n_i k_\mathrm{B} T \ln{\left(\frac{h \omega_i}{2 \pi k_\mathrm{B} T}\right)}} ,
\end{equation}
where $k_\mathrm{B}$, $h$, and $T$ mean the Boltzmann constant, Planck constant, and temperature, respectively, and $n_i$ and ${\omega}_i$ correspond to the number of atoms and angular frequency of Einstein oscillators of atomic species $i$, respectively. We use a spring constant of 5 eV/\AA$^2$ throughout this work regardless of atomic species.

For the liquid phase, we select for the reference system the “cut and shifted” LJ potential~\cite{lj1}. In Ref.~\cite{lj1}, the residual free energy of the LJ fluid in reference to the ideal gas was parameterized into an equation of state, which provides highly accurate free energies over a wide range of temperatures and densities. The free energy of the ideal gas is given by
\begin{equation}
	F = -k_\mathrm{B}T \sum_{i}{\ln{\left(\frac{V^{n_i}}{\Lambda_i^{3n_i}n_i!}\right)}} ,
\end{equation}
\begin{equation}	
	\Lambda_i = \frac{h}{\sqrt{2\pi m_i k_\mathrm{B}T}} ,
\end{equation}
where $V$ is the volume of the system and $\Lambda_i$ is the thermal De Broglie wavelength of the atomic species $i$ with the atomic mass of $m_i$. To avoid a phase transition along the integration path, the depth of the LJ potential is controlled such that the LJ fluid becomes supercritical, and the diameter of LJ particles is chosen to have a nearest-neighbor distance similar to the final state~\cite{ti2}.

\subsection{Semigrand ensemble simulations}
Taking the example of a binary system made of atoms A and B, the difference of chemical potentials is written as
\begin{equation}
	\label{g_vs_mu} \Delta\mu(x,T) \equiv \mu_\mathrm{B}(x,T) - \mu_\mathrm{A}(x,T) = \frac{\partial G(x,T)}{\partial x} ,
\end{equation}
where $x$ is the mole fraction of species B and $G$ is the Gibbs free energy. $\Delta\mu(x,T)$ can be obtained by the semigrand ensemble, a subset of the grand-canonical ensemble in which the number of atoms is fixed but chemical identities can change freely~\cite{sgc0,sgc1,sgc2,mcswap1}. In practice, the equilibration within the semigrand ensemble is achieved by hybridizing MD simulations with Monte Carlo (MC) swap of atomic species. The MC particle swap is accepted by the Metropolis criterion defined as
\begin{equation}
	p = \min{\left[ 1, \exp{\left(-\frac{\Delta E - \Delta\mu N\Delta x}{k_\mathrm{B}T}\right)}\right ]} ,
\end{equation}
where $\Delta E$ and $\Delta x$ indicate the change of energy and composition of the simulation cell due to the test flipping of atomic species respectively, and $N$ is the total number of atoms~\cite{mcswap1}. After sufficient MD-MC runs, the equilibrium composition $x$ is obtained for the given $\Delta\mu$. By iterating the semigrand ensemble simulations over a range of $\Delta\mu$, $x(\Delta\mu)$ and its inverse $\Delta\mu(x)$ are obtained at given $T$, and the free energy $G(x,T)$ is obtained in turn by integrating Eq.~\ref{g_vs_mu}. A more formal derivation~\cite{sgc0}, practical implementation~\cite{mcswap1}, and application examples~\cite{sgc1,sgc2} of the semigrand ensemble are referred to the literature. During the MD simulations, the isobaric condition is imposed to consider composition-dependent of lattice parameters. 

\section{RESULTS AND DISCUSSIONS}
\label{secresult}
\subsection{NNP training}
\label{sec3a}
The DFT data set for training NNPs consists of pure phases, solid solutions, and their melts. For pure phases of MgO and CaO, the data set first contains rocksalt crystals under volume-conserving uniaxial, hydrostatic, or shear strain, whose ranges are -5\% to 5\%, -2\% to 4\%, and -5\% to 5\%, respectively. For intermediate compositions, we generate 100 random alloys in the rocksalt structure (Mg$_x$Ca$_{1–x}$O) containing 100 atoms in $x=0.08,0.2,0.8,0.92$. For each composition, the lattice parameter is obtained by relaxing the cell shape and volume. To sample thermal vibrations of solids as well as liquid phases, the crystals of pure phases and random alloys (the most and least stable configurations among the 100 structures) are heated from 300 K to 2000, 4000, 6000, and 8000 K with a duration time of 1 ps at each temperature. Two independent MD simulations are performed in constant pressure (NPT) or constant volume (NVT) ensembles, where temperatures are modulated with the Langevin \cite{langevin1} or Nos{\'e}-Hoover \cite{nosehoover1} thermostats, respectively. We note that both ensembles are complementary in constructing data sets; while the NPT data set includes the thermal expansion of solid and liquid phases, NVT data set contains interactions between atoms at short distances, which helps prevent short-bond failures of NNPs during MD simulations along the thermodynamic integration path. We find that the pure phases and random alloys melt at 8000 and 6000 K, respectively. By including these melting processes, NNP may learn the interface between the solid and liquid phases required for coexistence simulations. Those MD trajectories are sampled with the interval of 40 and 10 fs at 300--4000 K and 6000--8000 K, respectively, and included in the data set after accurate single-shot DFT calculations. The whole data set contains 5,670 structures and 552,096 atoms for PBE and SCAN respectively. (See the Supplementary Materials for details.)  

\begin{figure}
	\includegraphics{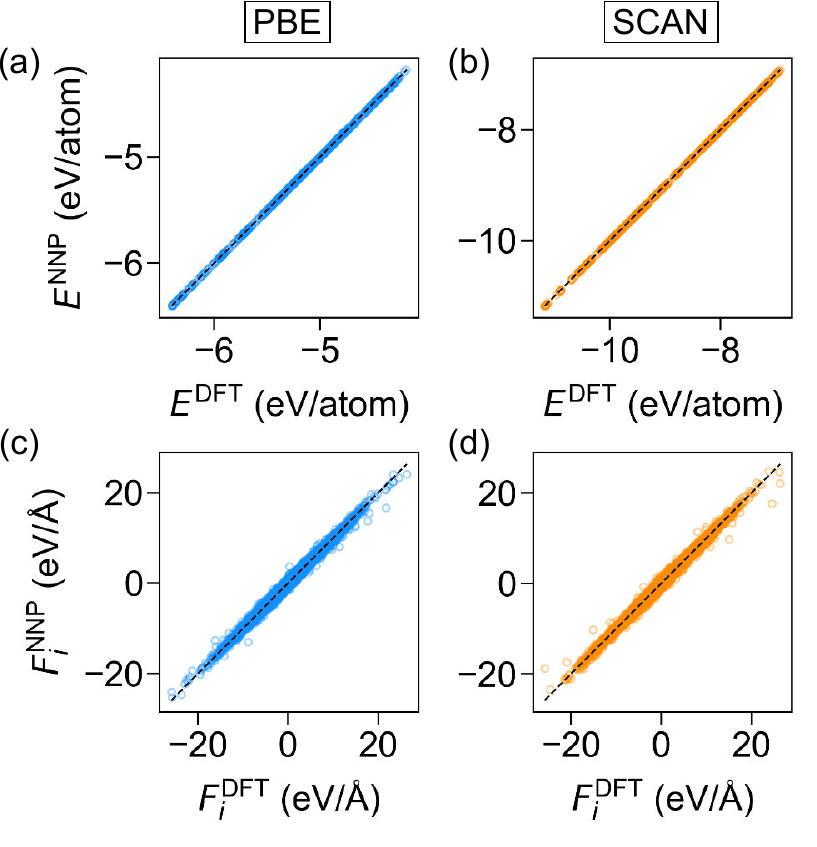} 
	\caption{\label{parityplot} Parity plots between DFT and NNPs, comparing energies ($E$) ((a) and (b)) and force component ($F_i, i = x, y, z$) in the Cartesian coordinate ((c) and (d)) for validation sets. The functional used for the reference data set is shown at the top.}
\end{figure}

We generate single NNP for PBE and SCAN functionals, named as PBE-NNP and SCAN-NNP, respectively, which is used for the whole calculations.
10\% of the data is randomly selected as a validation set to monitor overfitting. The root mean square error (RMSE) of NNPs on the training and validation set is presented in Table~\ref{rmse}, indicating that the accuracy of NNPs is satisfactory. 
The parity plots in Fig.~\ref{parityplot} display correlations of the energy and force components between DFT and NNP for the validation sets, showing that both PBE-NNP and SCAN-NNP well reproduce the reference DFT results. The slightly higher energy RMSE of SCAN-NNP (Table~\ref{rmse}) could be attributed to a wider energy range of the data set as seen in Figs.~\ref{parityplot}(a) and (b). In comparison, the force RMSE between the two NNPs is comparable since the magnitude of the force is similar in both data sets (see Figs.~\ref{parityplot}(c) and (d)).

\begin{table}[!]
	\renewcommand{\arraystretch}{0.8}
	\setlength{\tabcolsep}{6pt}
	\caption{\label{rmse} The root mean square error (RMSE) for the energy and force on training (T) and validation (V) sets. PBE-NNP and SCAN-NNP represent NNPs that are trained with the corresponding functional. In averaging errors in the force, the three-dimensional Euclidean distance is measured between DFT and NNP forces.}
	\begin{ruledtabular}
	\begin{tabular}{ lcccc }
				& \multicolumn{2}{c}{Energy (meV/atom)} & \multicolumn{2}{c}{Force (eV/\AA)} \\
		    & T               & V              & T               & V                   \\  \hline
		PBE-NNP  & 4.0             & 4.1            & 0.24            & 0.28                \\
		SCAN-NNP & 5.1             & 5.5            & 0.24            & 0.29                \\
	\end{tabular}
	\end{ruledtabular}
\end{table}

\subsection{Test of NNP on pure phases}
In Table~\ref{static}, the trained NNPs are further validated by comparing various properties of pure phases. We first compare structural and mechanical properties of rocksalt MgO and CaO at 0 K. It is seen that PBE overestimates the lattice parameters by 0.6--0.8\%, while SCAN underestimates by 0.3--0.6\%, in better agreements with experiment~\cite{mech1,mech2}. The elastic constants are also reproduced more accurately by SCAN than PBE, except for $C_{12}$. It is seen that each NNP well reproduces corresponding DFT results - lattice parameters within 0.001 \AA{} and elastic constants within 16.7\% (largest for off-diagonal component $C_{12}$ in SCAN-NNP).

\begin{table*}
	\caption{\label{static} Equilibrium lattice parameter ($a_0$), bulk modulus ($B$), and elastic constants ($C_{11}$, $C_{12}$ and $C_{44}$) of rocksalt MgO and CaO at 0 K. The properties are calculated after cell relaxations, and elastic constants are calculated by applying strains smaller than 0.5\%. Relative errors are presented in the parentheses with respect to the experiments (for DFT) or DFT calculations (for NNP), respectively.}
	\begin{ruledtabular}
	\begin{tabular}{llccccc}
			& Property & PBE& PBE-NNP & SCAN& SCAN-NNP & Experiment \\  \hline										
MgO	&	$a_0$ (\AA)	&	4.246 (0.8\%)	&	4.247 (0.0\%)	&	4.186 (-0.6\%)	&	4.186 (0.0\%)	&	4.213\footnote{Reference~\cite{mech1}.}	\\
&	$B$ (GPa)	&	153.3 (-6.9\%)	&	158.6 (3.4\%)	&	174.4 (5.9\%)	&	170.9 (-2.0\%)	&	164.7\footnote{Reference~\cite{mech2}.}	\\
&	$C_{11}$ (GPa)	&	273.8 (-10.7\%)	&	281.4 (2.8\%)	&	327.1 (6.6\%)	&	317.6 (-2.9\%)	&	306.7\footnotemark[2]	\\
&	$C_{12}$ (GPa)	&	93.1 (-0.6\%)	&	97.3 (4.5\%)	&	98.0 (4.6\%)	&	97.6 (-0.5\%)	&	93.7\footnotemark[2]	\\
&	$C_{44}$ (GPa)	&	145.2 (-7.9\%)	&	132.4 (-8.8\%)	&	160.8 (2.0\%)	&	146.0 (-9.2\%)	&	157.6\footnotemark[2]	\\ [1ex]

CaO	&	$a_0$ (\AA)	&	4.839 (0.6\%)	&	4.840 (0.0\%)	&	4.797 (-0.3\%)	&	4.797 (0.0\%)	&	4.811\footnotemark[1]	\\
&	$B$ (GPa)	&	105.2 (-7.7\%)	&	105.5 (0.3\%)	&	115.8 (1.6\%)	&	118.5 (2.4\%)	&	114.0\footnotemark[1]	\\
&	$C_{11}$ (GPa)	&	203.1 (-9.0\%)	&	203.6 (0.2\%)	&	241.7 (8.2\%)	&	232.3 (-3.9\%)	&	223.3\footnotemark[1]	\\
&	$C_{12}$ (GPa)	&	56.3 (-5.0\%)	&	56.5 (0.3\%)	&	52.8 (-10.9\%)	&	61.7 (16.7\%)	&	59.3\footnotemark[1]	\\
&	$C_{44}$ (GPa)	&	74.8 (-7.7\%)	&	70.7 (-5.4\%)	&	86.0 (6.2\%)	&	76.3 (-11.3\%)	&	81.0\footnotemark[1]	\\
	\end{tabular}
	\end{ruledtabular}
\end{table*}

\begin{figure}
	\includegraphics{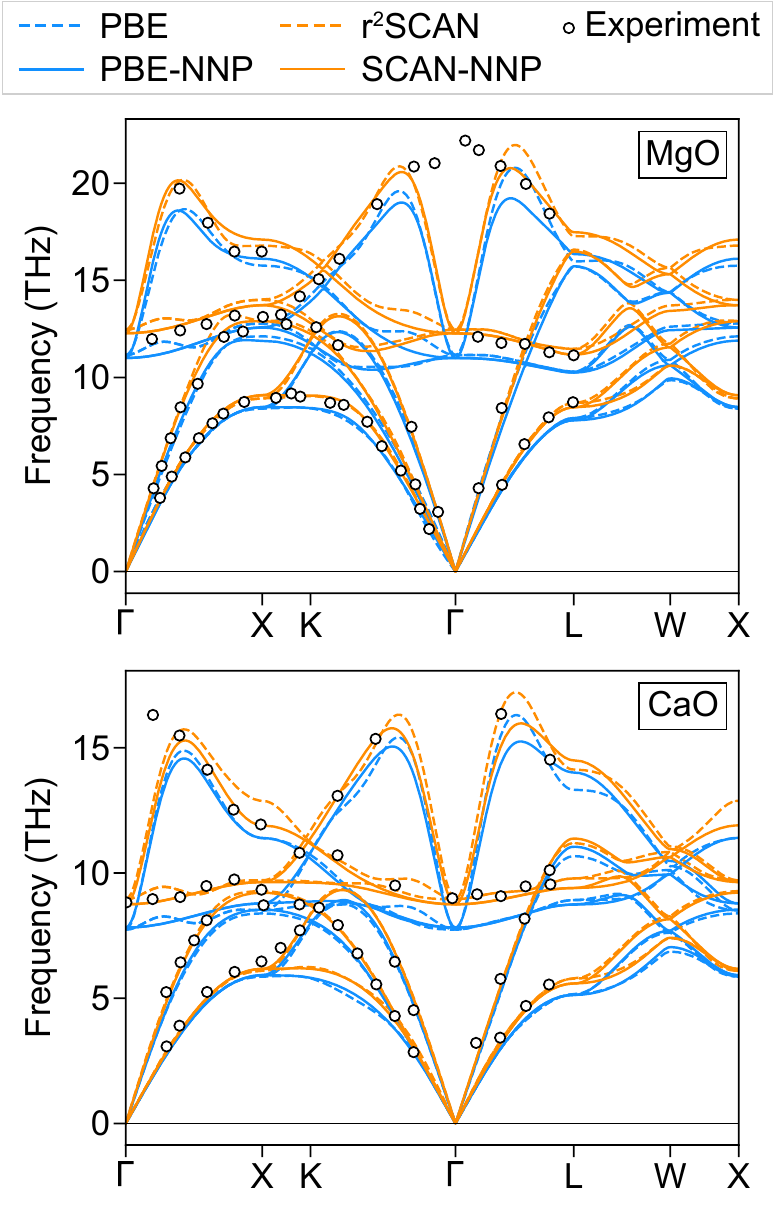}
	\caption{\label{phonon} Phonon dispersion along the high-symmetry points of the rocksalt phase of MgO and CaO. We use the r$^2$SCAN functional instead of the original SCAN in this case for better numerical convergence of lattice dynamics~\cite{scan4}. Experimental measurements are adopted from Refs.~\cite{ph1} (MgO) and~\cite{ph2} (CaO).}
\end{figure}

In Fig.~\ref{phonon}, we compute phonon dispersions and compare them with experiments. The phonon dispersions are calculated using the \texttt{Phonopy} code~\cite{ph4} with the finite displacement method and a 5$\times$5$\times$5 repetition of the primitive cell. In Ref.~\cite{scan4}, it was tricky to obtain phonon dispersions with the SCAN functional due to unstable convergences, which is also confirmed in the present work as the phonon dispersions calculated within the SCAN functional exhibit spurious imaginary modes for rocksalt MgO and CaO. Instead, we employ r$^2$SCAN functinoal~\cite{scan3} for phonon calculations, as it exhibits better numerical convergences while maintaining the accuracy of the original SCAN. As shown in Fig.~\ref{phonon}, the r$^2$SCAN functional accurately reproduces the lattice dynamics of the experiments~\cite{ph1,ph2}. On the other hand, PBE calculations underestimate the phonon frequencies. We do not consider the modifications of optical branches due to the long-range Coulomb interactions (LO-TO splitting), resulting in the deviations of optical branches near the $\Gamma$ point. In Fig.~\ref{phonon}, NNPs successfully reproduce the phonon dispersions by DFT regardless of the functional type.

\begin{figure}
	\includegraphics{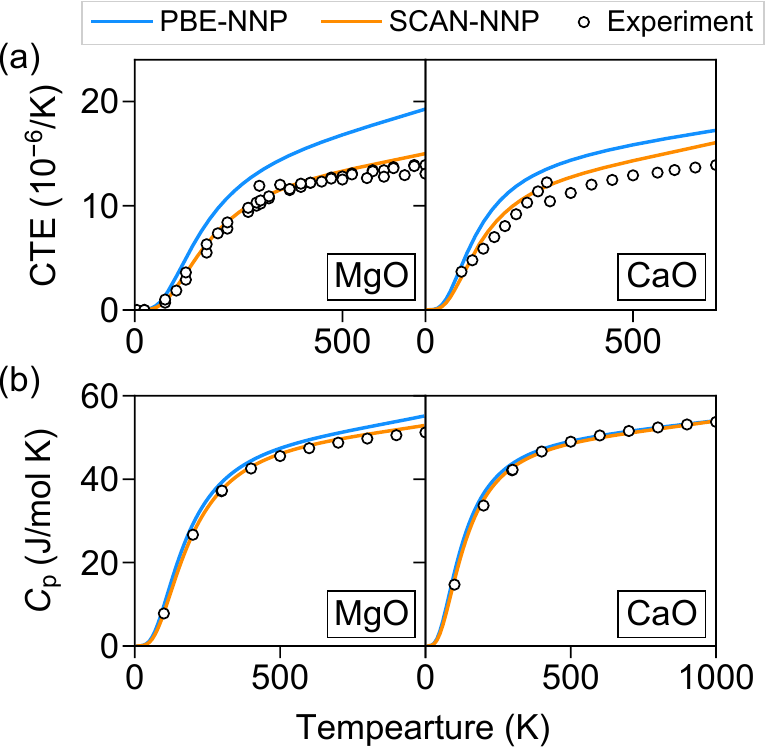}
	\caption{\label{thermal} Thermal properties of MgO and CaO calculated by quasi-harmonic approximations: (a) linear coefficient of thermal expansion (CTE) and (b) heat capacity at constant pressure ($C_\mathrm{p}$). Experimental values of CTE for MgO and CaO are from Ref.~\cite{cte1,cte2,cte3} and Ref.~\cite{cte3,cte4}, respectively, while $C_\mathrm{p}$ is from the thermochemical tables~\cite{cp1}.}
\end{figure}

To benchmark thermal properties of solids at constant pressures, the linear coefficient of thermal expansion (CTE) and heat capacity ($C_{\rm p}$) are calculated in Fig.~\ref{thermal} within the quasi-harmonic approximation~\cite{ph4}. As can be seen in Fig.~\ref{thermal}(a), for both pure phases, the predicted CTE in SCAN-NNP compares favorably to the experiments, whereas PBE-NNP overestimates it by about 20-30\%. Similarly, Fig.~\ref{thermal}(b) shows that $C_{\rm p}$ of MgO agrees well between SCAN-NNP and experiment, while PBE-NNP slightly overestimates it. For CaO, $C_{\rm p}$ is accurately predicted by both NNPs, although SCAN-NNP and PBE-NNP perform slightly better at temperatures below and above 350 K, respectively. 

\begin{figure*}
	\includegraphics{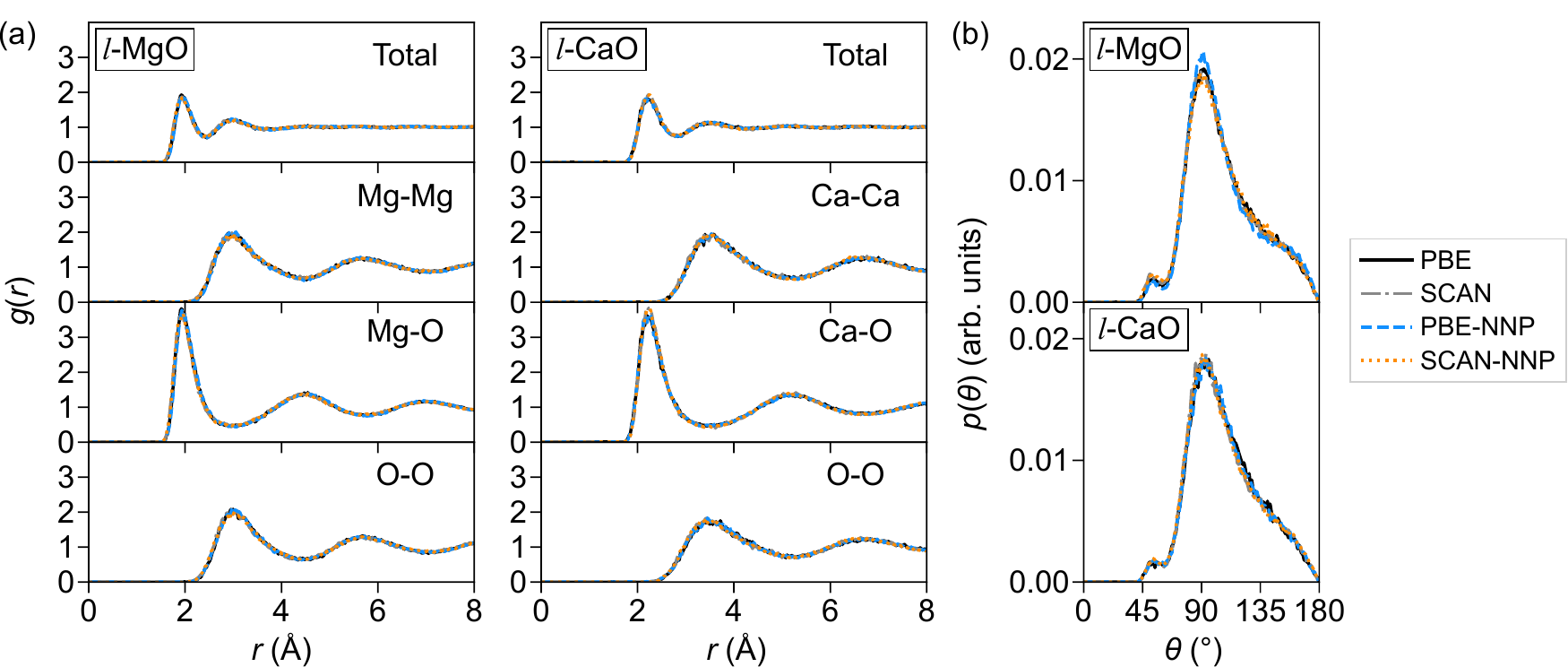}
	\caption{\label{rdfadf} (a) Total and partial radial distribution functions ($g(r)$) and (b) total angular distribution functions ($p(\theta)$) of liquid MgO and CaO. }
\end{figure*}

We next compare structural properties of the liquid phases, which are obtained by employing 100-atom supercells and NVT ensembles with temperatures of 3100 and 2850 K for MgO and CaO, respectively. The radial and angular distribution functions (RDF and ADF, respectively) are averaged over 40-ps MD simulations, preceded by 5-ps pre-melting at twice the temperature and 10-ps equilibration. The total and atom-resolved RDFs in Fig.~\ref{rdfadf}(a) indicate that the first peaks are dominated by heteropolar pairs for both liquid MgO and CaO ($l$-MgO and $l$-CaO, respectively). The first peaks lie at 2.0 and 2.2 \AA{} for MgO and CaO respectively, where the difference stems from the larger ionic radius of Ca than that of Mg. 
The second peaks consist of mostly homopolar pairs, with similar distributions among the pairs. 
The ADFs are shown in Fig.~\ref{rdfadf}(b), and both phases commonly exhibit a major peak at 90\textdegree{} and shoulder peaks around 50\textdegree{} and 150\textdegree{}. Both NNPs well reproduce main features in the RDF and ADF from DFT calculations. It is noticeable that the liquid structures of PBE-NNP and SCAN-NNP are hardly distinguishable despite the significant differences in the solid phase.

\subsection{Test of NNP on pseudo-binary mixtures}
\label{sec3c}
In this subsection, we test the accuracy of NNPs for solids and liquids at intermediate compositions. To this end, we first compare the formation energies of substitutional defects in solids that affect the free energy of mixing at low concentrations. The defect formation energy ($D_\mathrm{f}$) is defined as follows:
\begin{equation}
	\label{eq_defect} D_\mathrm{f} = E_{\mathrm{defect}} - \sum_{i}{N_i E_i} ,
\end{equation}
where $E_{\mathrm{defect}}$ means the total energy of the supercell containing a point defect, and $N_i$ and $E_i$ ($i=\mathrm{MgO, CaO}$) indicate the number of formula unit in the supercell and the energy of pure phases, respectively. As can be seen from Table~\ref{substitution}, NNPs  reproduce DFT formation energies of the substitutional defects within 3\%. Both PBE and SCAN produce a larger $D_{\rm f}$ for Ca$_{\rm Mg}$ than for Mg$_{\rm Ca}$, which implies a lower solubility of the former. It is also seen that SCAN produces a higher $D_{\rm f}$ than PBE by 0.2 eV, which affects the solubility limit as will be shown later. 

\begin{table}
	\caption{\label{substitution} The formation energy of substitutional defects in eV. $\mathrm{Ca_{Mg}}$ and $\mathrm{Mg_{Ca}}$ mean a single-atom impurity of CaO in MgO and MgO in CaO respectively, where we use 216-atom supercells to evaluate the formation energy of point defects.}
	\begin{ruledtabular}
	\begin{tabular}{l cccc}
		Type &	PBE&	PBE-NNP&	SCAN&	SCAN-NNP \\  \hline
		$\mathrm{Ca_{Mg}}$ &	1.00& 	0.98& 	1.21& 	1.19 \\
		$\mathrm{Mg_{Ca}}$ &	0.67& 	0.70& 	0.86& 	0.82 \\
	\end{tabular}
	\end{ruledtabular}
\end{table}

\begin{figure}
	\includegraphics{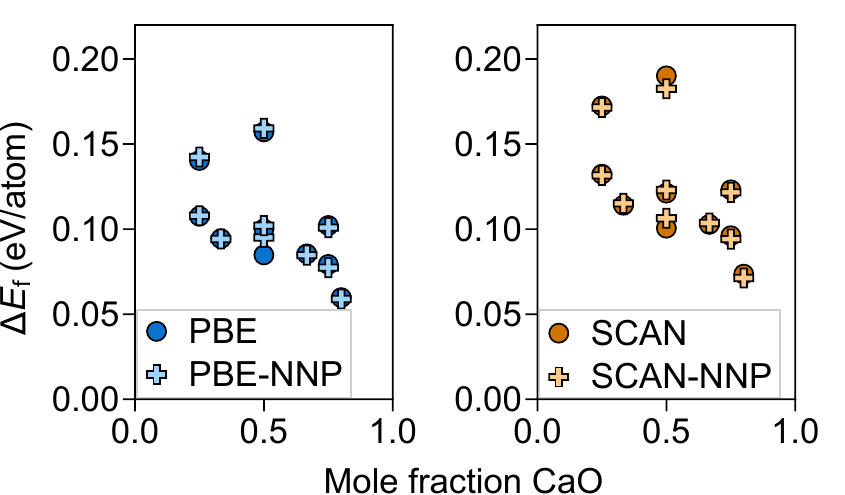}
	\caption{\label{ordered} Formation energy ($\Delta E_{\rm f}$) of ten ordered structures evaluated within PBE or SCAN functional and corresponding NNPs.}
\end{figure}

Next, we compare the formation energies of ordered structures at intermediate compositions. We consider ten ordered structures~\cite{mgocao2} by exchanging cations in the rocksalt lattice, including $L1_0$, $L1_1$, NbP, Ni$_4$Mo, $L1_2$, $D0_{22}$, and MoPt$_2$ structures where the latter three structures include both Mg- and Ca-rich stoichiometries. The formation energy per atom ($\Delta E_\mathrm{f}$) is defined as follows:
\begin{equation}
	\label{eq_ordered} \Delta E_\mathrm{f} = \frac{1}{2\sum_{i}{N_i}} \left[ E_{\mathrm{SC}} - \sum_{i}{N_i E_i} \right],
\end{equation}
where $E_{\mathrm{SC}}$ means the total energy of the ordered structure and other notations are the same as in Eq.~\ref{eq_defect}. 
The $\Delta E_\mathrm{f}$'s computed by DFT and NNPs are compared in 
Fig.~\ref{ordered}, showing that NNPs closely reproduce corresponding DFT results within 10 meV/atom. It is understandable that the errors in $\Delta E_\mathrm{f}$ are maximum at Mg$_{0.5}$Ca$_{0.5}$O, as the training set consists of pure phases and mixtures of up to 20\% mole fractions. It is seen that none of the ordered phases are energetically favorable with respect to the pure phases, with $\Delta E_\mathrm{f}$ greater than 50 meV/atom. We also note that the magnitude of $\Delta E_\mathrm{f}$ is larger in SCAN than PBE, which is consistent with $D_{\mathrm f}$.

\subsection{Free energy of pure phases}
\label{sec3d}
With the accuracy on solid and liquid phases confirmed, the trained NNPs are used in the thermodynamic integration to calculate free energies of the solid and liquid phases. We employ a 10-point Gauss-Legendre quadrature to evaluate the integral in Eq.~\ref{eq_ti} using the lattice parameters obtained from NPT simulations at zero pressure. For all the phases, each point in the quadruture is evaluated by employing a 1,000-atom supercell and 2-ps equilibration followed by 5-ps sampling for the temporal average. We use the Langevin thermostat~\cite{langevin1} with the center of mass fixed to avoid drift of the atoms~\cite{ti1,ti5,ti6}. To determine convergence, we use the block standard error (BSE) as a measure of uncertainty~\cite{bse1}.

\begin{figure}
	\includegraphics{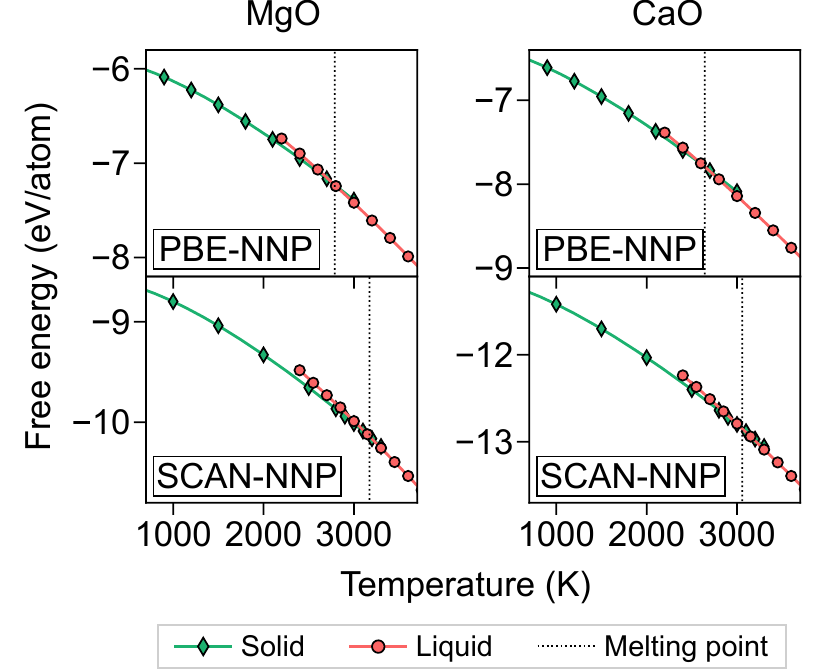}
	\caption{\label{freecurve} Free energy curves of rocksalt and liquid phases of MgO and CaO. Free energies that are directly calculated from the thermodynamic integration are represented by diamonds (solid) and disks (liquid), and the free energy models are shown in solid lines. Melting points are indicated by vertical dotted lines at the intersection of the free energy curves.}
\end{figure}

Figure~\ref{freecurve} shows the computed free energies of pure phases, which are fitted to an analytical free energy model as follows:
\begin{equation}
	\label{fpure}
	G(T) = a + bT + cT \ln{T} + dT^2 + eT^{-1} ,
\end{equation}
where $a$, $b$, $c$, $d$, and $e$ are fitting parameters. Similar function forms were used in the previous thermodynamic calculations~\cite{mgocao10} and MD studies~\cite{ti4,ti5}. The error of fit is less than 2 meV/atom in both solid and liquid phases, which is on the order of the BSE of each point and sufficient to obtain melting properties. By fitting to the smooth function in Eq.~\ref{fpure}, the determination of temperature-dependent free energies becomes robust against statistical fluctuations in the numerical integration. We add that the specific function form has negligible effects on the melting properties as long as the free energy data are well fitted into the model.  

The resulting free-energy curves of MgO and CaO are shown in Fig.~\ref{freecurve} as solid lines, and melting properties obtained from intersections of the curves are summarized in Table~\ref{meltingprop}. The melting point of MgO is calculated as 2787 K by PBE-NNP, which is consistent with the previous works at the PBE level, 2747 K by DFT calculations~\cite{mgocao5} and 2698 K by the Gaussian approximation potential (GAP)~\cite{mgocao6}. However, these values are significantly underestimated compared to the experimental range of 3040--3250 K~\cite{mgocao5,mgocao10}. In contrast, the SCAN-NNP produces an improved melting point of 3173 K, which is within the experimental range and agrees reasonably with the previous SCAN-DFT calculations (3032 K) or SCAN-GAP (3072 K). The entropy of fusion and slope of melting curve of MgO are mostly consistent among the same functional. On the other hand, the melting point of CaO is computed to be 2640 K by PBE-NNP, which is far below the experimental data of 2850--3220 K~\cite{mgocao10}. The SCAN-NNP better predicts the melting point of CaO to be 3057 K, which is within the experimental range.

\begin{table*}
	\caption{\label{meltingprop} Melting point ($T_\mathrm{m}$), entropy of fusion ($\Delta S_\mathrm{m}$), and slope of the melting curve ($dT/dP$) of MgO and CaO. PBE-GAP and SCAN-GAP represent the Gaussian approximation potential (GAP) trained by the PBE and SCAN data sets, respectively. Approaches refer to thermodynamic integration (TI), coexistence method (Coexist), and interface pinning method with a correction by thermodynamic perturbation theory (IP+TP)~\cite{ip1}. The block standard errors are also provided. Results are from this work unless references are given.}
	\begin{ruledtabular}
	\begin{tabular}{ll cccc cccc}
		& Property & PBE-NNP & PBE-NNP & SCAN-NNP & SCAN-NNP & PBE\footnote{Reference~\cite{mgocao5}.} & SCAN\footnotemark[1]
					& PBE-GAP\footnote{Reference~\cite{mgocao6}.} & SCAN-GAP\footnotemark[2] \\  \hline
		MgO& $T_\mathrm{m}$ (K)& 2787±30&	2786±1.5&	3173±33&	3181±1.7&	2747±59&	3032±53&	2698±23&	3072±25 \\
		   &	$\Delta S_\mathrm{m}$ ($k_\mathrm{B}$/atom)&	1.61& 	-&	1.58 &	-&	1.63 &	1.70 &	1.57& 	1.50 \\
		   &   	$dT/dP$ (K/GPa)&	150&	-&	140&	-&	155&	134&	153&	140	\\		
		   &	Method&	TI&	Coexist&	TI&	Coexist&	TI&	TI&	IP+TP&	IP+TP \\ [1ex]
		CaO&	$T_\mathrm{m}$ (K)&	2640±30&	2659±1.5&	3057±35&	3097±2.0& &&& \\
		   & $\Delta S_\mathrm{m}$ ($k_\mathrm{B}$/atom)&	1.65& 	-&	1.56& 	-& &&& \\
		   & $dT/dP$ (K/GPa)&	181&	-&	156&	-& &&& \\
		   & Method&	TI&	Coexist&	TI&	Coexist& &&& \\
	\end{tabular}
	\end{ruledtabular}
\end{table*}

For a further check, the melting points of the pure phases are recalculated with the coexistence method. In this method, the simulation cell contains solid and liquid phases and the interfaces between them, which is directly equilibrated to identify the transition temperature at which the interface stops moving. We employ a 16,000-atom simulation cell that is a 10$\times$10$\times$20 replication of the conventional unit cell. The initial simulation cell is prepared in NPT ensembles, with the initial guess on melting points calculated from the thermodynamic integration. Half of the simulation cell is melt-quenched to the tentative melting point while the other atoms are frozen. Then the simulation cell is equilibrated within the NPH ensemble for 100 ps, and the temperature is sampled for another 100 ps. When we test the cell size effect with 2,000-atom simulation cells, the melting point shifts only by 6 K. As can be seen in Table~\ref{meltingprop}, the melting points calculated by the thermodynamic integration and coexistence methods agree within 40 K. 

\subsection{Phase diagram}
To construct the full phase diagram, we compute the free energies in semigrand ensembles at intermediate compositions. The isobaric ensemble is used to allow for the volume to change according to the composition during the MD simulations, and the cell size is the same as in the thermodynamic integration. The ensemble is equilibrated and sampled during 50,000 steps with the 2-fs time step, and attempts to swap between Mg and Ca atoms are set at 1\% of the cations per time step. Single run of the semigrand ensemble simulation at given $\Delta\mu$ and $T$ provides the corresponding equilibrium composition $x$. After carrying out the semigrand ensemble simulations over a set of $(\Delta\mu,T)$, one can obtain composition-dependent Gibbs free energies following the relation in Eq.~\ref{g_vs_mu}. 

\begin{figure*}
	\includegraphics{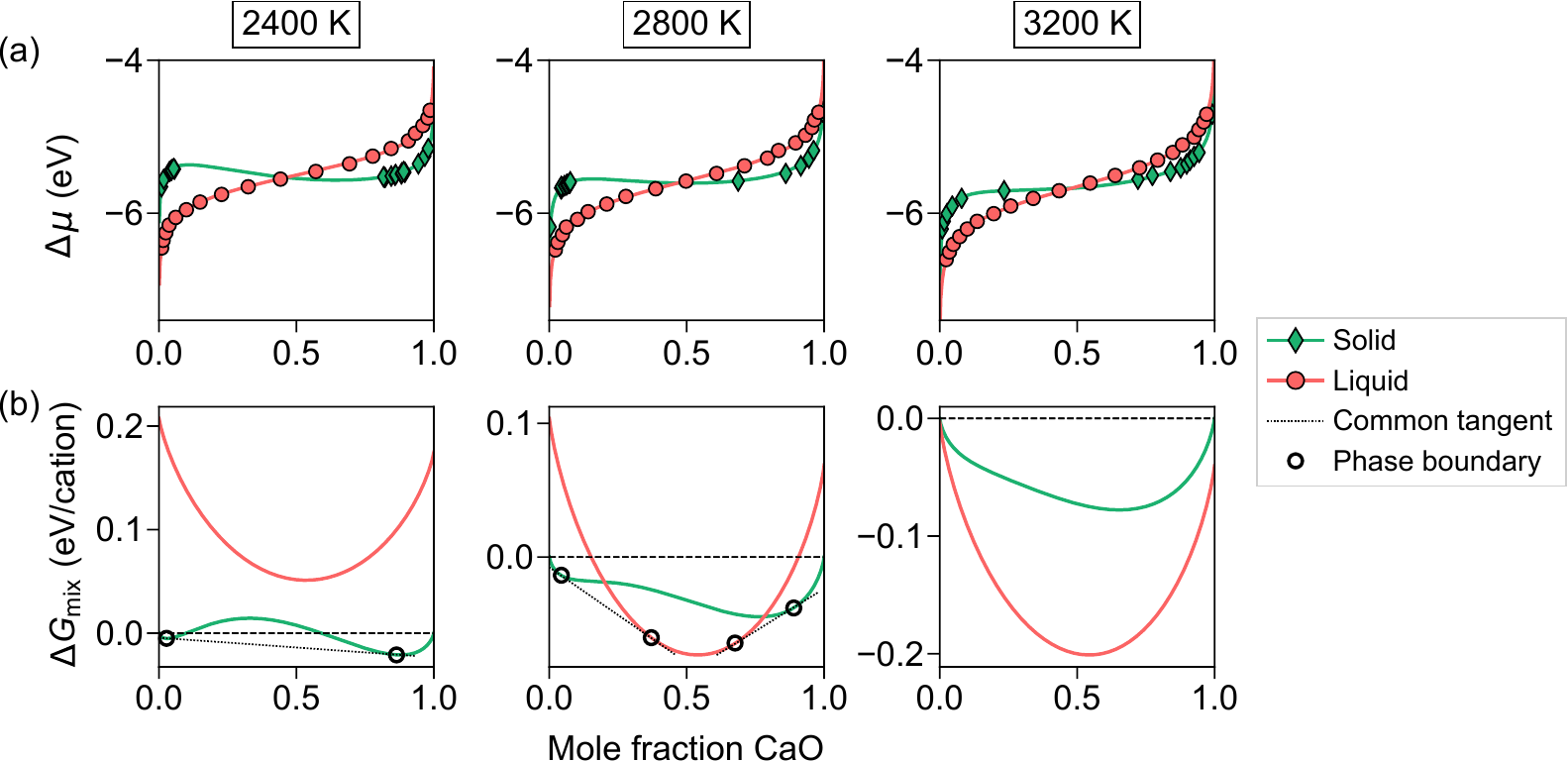}
	\caption{\label{semigrand} (a) Chemical potential difference ($\Delta \mu = \mu_{\mathrm{CaO}}-\mu_{\mathrm{MgO}}$) and (b) the free energy of mixing ($\Delta G_{\mathrm{mix}}$) as a function of composition at selected temperatures, which are calculated with SCAN-NNP. $\Delta G_{\mathrm{mix}}$ is defined as the difference from the free energy of pure solids. Symbols, solid lines, and dotted lines represent data points obtained from the semigrand ensemble simulations, the fitted free energy models, and the common tangents that determine phase boundary, respectively. Empty circles correspond to the phase boundary at the given temperature.}
\end{figure*}

In Fig.~\ref{semigrand}(a), results from the semigrand simulation using SCAN-NNP are shown for $\Delta \mu = \mu_{\rm CaO} - \mu_{\rm MgO}$. With solid solutions at 2400 K, there exists a $\Delta \mu$ range where the equilibrium composition is not unique due to the dependence on the initial composition. Because of this hysteresis, pure phases of MgO or CaO should be used as initial configurations to scan over end compositions. This is the reason why data points are empty for a range of intermediate compositions at 2400 K (and also 2800 K). The hysteresis weakens with the increasing temperature and almost disappears at 3200 K. For the liquid phase, such hysteresis does not exist at any simulation temperature.

The semigrand simulations are carried out for solid and liquid phases at least five temperatures spanning relevant domains in the phase diagram. (For example, in the case of SCAN-NNP, the simulation temperatures for solids (liquids) are sampled from 1200 (2400) K to 3200 (3300) K with the interval of 400 (100) K.) 
In order to interpolate free energies over the whole phase diagram and obtain $G(x,T)$ via integration of $\Delta\mu(x,T)$ following Eq.~\ref{g_vs_mu}, we introduce analytical models for the free energy~\cite{sgc2,mgocao10} and fit them to the simulation data in Fig.~\ref{semigrand}(a). First, the free energy is written as follows:
\begin{equation}
	\label{gtot} G(x,T) = \overline{G}(x,T) + \Delta G_{\mathrm{mix}}(x,T) ,
\end{equation}
where $x$ is the mole fraction of CaO and $\overline{G}(x,T)$ means the weighted average of free energies of pure phases: 
\begin{equation}
	\label{gbar} \overline{G}(x,T) = xG_{\mathrm{CaO}}(T) + (1-x)G_{\mathrm{MgO}}(T) ,
\end{equation}
where $G_{\mathrm{MgO}}$ and $G_{\mathrm{CaO}}$ are free energies of the pure phases  obtained in the previous subsection. In Eq.~\ref{gtot}, $\Delta G_{\mathrm{mix}}(x,T)$ means the residual free energy of mixing defined as
\begin{eqnarray}
	\label{dgmix} \Delta G_{\mathrm{mix}}(x,T) = k_{\mathrm{B}}T[x\ln{x} + (1-x)\ln{(1-x)}] \nonumber \\
	+ x (1-x) (A+Bx+Cx^2) ,
\end{eqnarray}
where the first term corresponds to the ideal free energy of mixing, and the second term reflects the non-ideality with the temperature-dependent parameters $A$, $B$, and $C$. The chemical potential model is derived from the relation in Eq.~\ref{g_vs_mu}, written as
	\begin{eqnarray} \label{dmu}
	&& \Delta \mu (x,T) = \mu_{\mathrm{CaO}} - \mu_{\mathrm{MgO}} \nonumber = \frac{\partial G(x,T)}{\partial x} \\
	&& = G_{\mathrm{CaO}}(T) - G_{\mathrm{MgO}}(T) + k_{\mathrm{B}}T \ln{\left(\frac{x}{1-x}\right)} \nonumber \\
	&& + A + 2(B-A)x + 3(C-B)x^2 -4Cx^3.
\end{eqnarray}
Equation~\ref{dmu} is fitted to the simulation data in Fig.~\ref{semigrand}(a), and the optimized models in solid lines are in good agreements with the simulation data. The parameters $A$, $B$, and $C$ are assumed to be linear with the temperature as in Ref.~\cite{mgocao10}, and the fitting RMSE of the solid phase is 5.1 and 7.3 meV/atom for PBE-NNP and SCAN-NNP, respectively, and the corresponding RMSEs in the liquid phase are 3.6 and 5.8 meV/atom, respectively.

Figure~\ref{semigrand}(b) shows the fitted $\Delta G_{\rm mix}$ in Eq.~\ref{dgmix} at the selected temperatures. At 2400 K, the free energy curve for the solid phase features a miscibility gap resulting from the two local minima at terminal solutions, while no liquid phase is thermodynamically stable throughout the composition. At the elevated temperature of 2800 K, the liquid phase becomes stable over a range of intermediate compositions, and so the eutectic point is expected to lie between 2400 K and 2800 K. Above 3200 K, the liquid phase is always stable over the solid phase as the temperature becomes higher than the melting point of both pure phases. The dotted lines in Fig.~\ref{semigrand}(b) are common tangents of stable phases, and the contacts are indicated by the circles. These contacts represent the phase boundary since the coexistence of those phases is thermodynamically favored over other compositions and phases.

\begin{figure*}
	\includegraphics{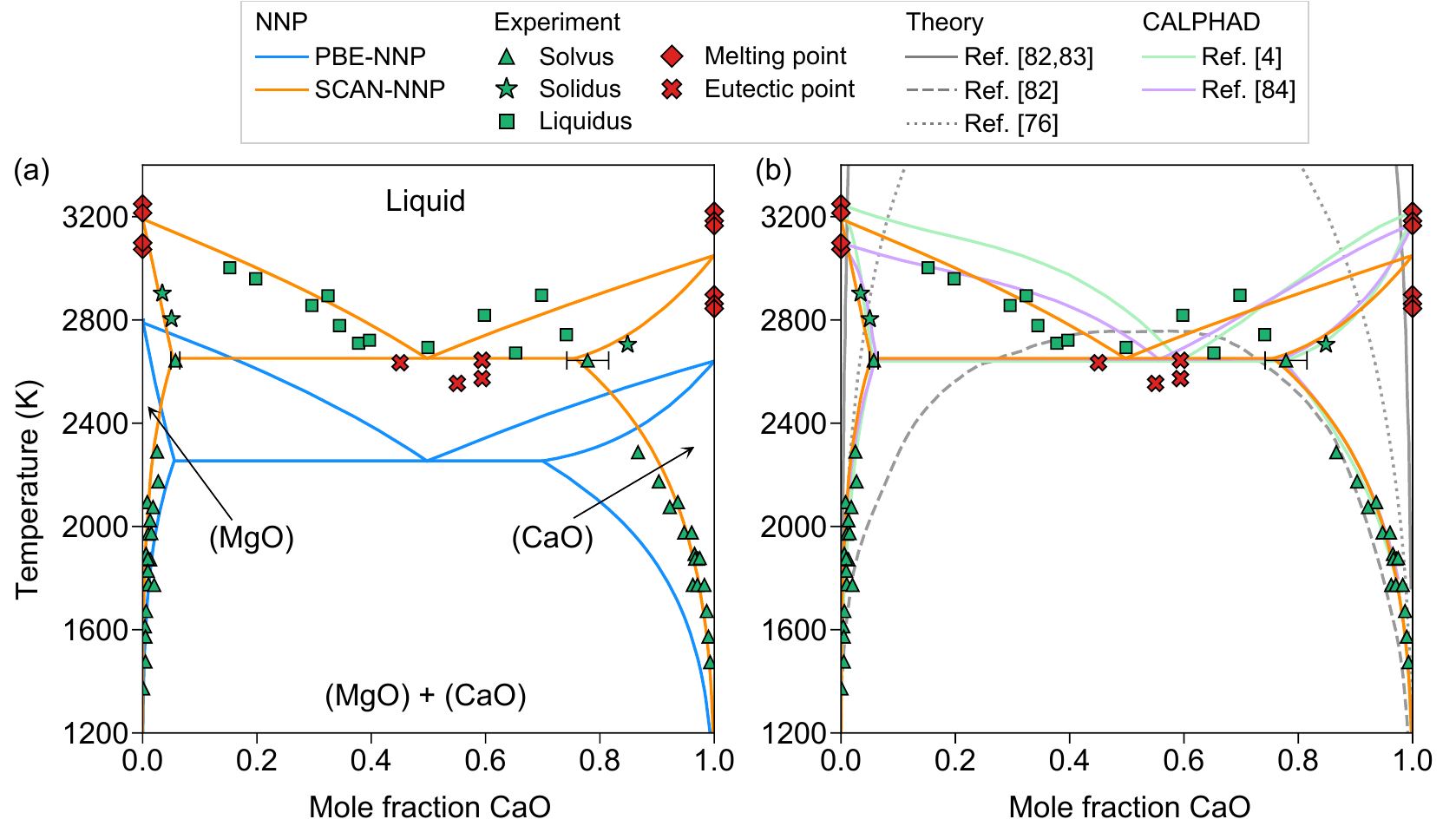}
	\caption{\label{phasediagram} Phase diagrams of MgO-CaO calculated with NNPs, compared to (a) experiments or (b) other theoretical works or CALPHAD modelings. Experimental solvus, solidus, and liquidus data are from Ref.~\cite{mgocao10} and the references therein. Theoretical solid-state phase diagrams that are calculated from classical potential (Ref.~\cite{mgocao1,mgocao11}), first-principles calculations with vibrational effects (Ref.~\cite{mgocao1}), or tight-binding calculations (Ref.~\cite{mgocao2}) are displayed in gray lines. The CALPHAD models are adopted from Ref.~\cite{mgocao10} and Ref.~\cite{mgocao12}.} 
\end{figure*}

The full phase diagrams constructed within NNPs are shown in Fig.~\ref{phasediagram}(a) together with experimental data. 
Based on the fitted analytical free energy models, we calculate the phase boundaries with the 1 K interval between 1200 and 3200 K. It is seen that both PBE-NNP and SCAN-NNP reproduce the characteristics of the MgO-CaO system such as eutectic points and solubility limits. In detail, the eutectic compositions predicted by PBE-NNP and SCAN-NNP are  0.50 and 0.49 for the mole fraction CaO, respectively, which are within the experimental observations of 0.45-0.60~\cite{mgocao10} (see red crosses). The eutectic temperature, on the other hand, is 2253 K and 2651 K for PBE-NNP and SCAN-NNP respectively, only the latter being close to the experimental range of 2550-2640 K. The failure of PBE-NNP is consistent with the underestimated melting points of the pure phases. The experimental solid solubility of CaO in MgO (MgO in CaO) at the eutectic temperature is 6\% (22\%) mole fraction CaO~\cite{eut4}, which are closely reproduced by SCAN-NNP within the error bar. The PBE-NNP can also reproduce the solid solubility of CaO in MgO at its own eutectic temperature, but the solubility of MgO in CaO is overestimated by about 10\%. The overestimation is related to the smaller formation energy of substitutional defects than with SCAN-NNP (see Sec.~\ref{sec3c}), which leads to thermodynamic preference toward mixing. Other experimental data regarding the solvus, solidus, and liquidus are all in good agreements with those by SCAN-NNP.

Figure~\ref{phasediagram}(b) compares the phase diagram by SCAN-NNP and those from other atomistic simulations (see gray lines). Previous theoretical works identified only solid-state phase diagrams of the MgO-CaO system with classical potentials~\cite{mgocao1,mgocao11} or first-principles calculations~\cite{mgocao1,mgocao2}. (To note, the effect of lattice vibration is considered only in Ref.~\cite{mgocao1}.)
It is seen that none of previous works produced correct solvus lines on both MgO- and CaO-rich sides.  
On the other hand, the results by CALculation of PHAse Diagrams (CALPHAD) modeling are also presented in Fig.~\ref{phasediagram}(b). While solvus lines are consistent with the SCAN-NNP results, eutectic point, solidus, and liquidus are at variance with each other, even among the CALPHAD data. This is because while solvus lines are validated through a number of experiments, the data for solidus and liquidus lines are sparse and scattered~\cite{mgocao10}. The mismatch of the phase boundaries from CALPHAD models is understandable because each model is fitted to different sets of data points. 

\section{CONCLUSION}
\label{secsummary}
We remark on the computational efficiency for constructing the phase diagram. The whole procedure, including the data set generation, NNP training, and free energy calculations with MD simulations, took about ten days of computing time on 400 cores of Intel\textsuperscript{\sffamily\textregistered} Xeon\textsuperscript{\sffamily\textregistered} Gold 6148 CPU running at 2.4 GHz. In detail, about five days were spent on generating data sets and training NNPs, and another five days on free energy calculations using NNPs and 1,000-atom cells. If identical free-energy calculations were carried out by purely DFT approaches, it would take several decades with the same computational resource, even assuming that the free energy calculations are done on smaller 200-atom simulation cells. This is mainly because the hybrid MC-MD simulations require a large amount of computing resources due to several million time steps.

In summary, we developed NNPs for the MgO-CaO pseudo-binary system and demonstrated construction of the full phase diagram. The accuracy of NNPs trained over PBE or SCAN data is confirmed by validation over diverse properties. Notably, SCAN-NNP outperformed PBE-NNP in most cases when compared with experiments. The full phase diagrams are determined from the free energy calculations employing thermodynamic integration and semigrand ensemble methods. Notably, SCAN-NNP produced a phase diagram that closely follows experimental measurements on liquidus, solidus, and solvus lines, including the eutectic point and solid solubility limits. 
In conclusion, we believe that this work will pave the way to the \textit{ab initio} CALPHAD approach with high prediction accuracies.

\begin{acknowledgments}
This work was supported by the Creative Materials Discovery Program (RIAM) through the National Research Foundation of Korea (NRF) funded by the Ministry of Science and ICT (2017M3D1A1040689). The computations were carried out at Korea Institute of Science and Technology Information (KISTI) National Supercomputing Center (KSC-2021-CRE-0476).
\end{acknowledgments}

\bibliography{ref}

\end{document}